\documentclass[fleqn,usenatbib]{mnras}

\usepackage{newtxtext,newtxmath}
\usepackage{graphicx}
\usepackage[T1]{fontenc}

\DeclareRobustCommand{\VAN}[3]{#2}
\let\VANthebibliography\thebibliography
\def\thebibliography{\DeclareRobustCommand{\VAN}[3]{##3}\VANthebibliography}

\usepackage{graphicx}	
\usepackage{amsmath}	

\title[Optical continuum variability of WLQs]{Weak-emission-line quasars: A new clue from their optical variability}

\author[Kumar et al.]{
Ritish Kumar,$^{1}$\thanks{E-mail: ritishshield@gmail.com}
Gopal-Krishna,$^{2}$
Hum Chand$^{1}$
and Vibhore Negi$^{3}$
\\
$^{1}$Department of Physics and Astronomical Science, Central University of Himachal Pradesh, Dharamshala, 176215, India\\
$^{2}$UM-DAE Centre for Excellence in Basic Sciences (CEBS), Vidyanagari, Mumbai - 400098, India\\
$^{3}$Inter-University Centre for Astronomy and Astrophysics, Post Bag 4, Ganeshkhind, Pune, 411007, India
}

\date{Accepted XXX. Received YYY; in original form ZZZ}

\pubyear{\the\year{}}

\begin{document}
\label{firstpage}
\pagerange{\pageref{firstpage}--\pageref{lastpage}}
\maketitle

\begin{abstract}
Weak-emission-line QSOs (WLQs) are an enigmatic subclass of the QSO population, as their optical/UV spectra are marked by abnormally weak (or absent) emission lines. To obtain much-needed additional clues to the origin of this and other known peculiarities of WLQs, we have determined the `ensemble' optical variability characteristics for a large, well-defined sample of 76 radio-quiet WLQs and also for a matched control sample comprising 603 normal radio-quiet QSOs. This analysis was done using their light-curves recorded in the $g$ and $r$ bands, under the ZTF survey during 2018-2024, with a typical cadence of 3 days. We find that, compared to normal QSOs, WLQs exhibit systematically milder optical variability on month/year-like time scales (by a factor of $\sim$ 1.76$\pm$0.05 in amplitude). We have independently verified this by carrying out an equivalent analysis of the V-band light-curves acquired under the CRTS during 2007- 2014, with a typical cadence of 10 days. This new observational differentiator between WLQs and normal QSOs may provide clues to understanding the intriguing nature of WLQs. It is proposed that the clumpiness of the torus material flowing into the central engine may play a key role in explaining the observed differences between the WLQs and normal QSOs.
\end{abstract}




\section{Introduction}
\label{introduction}
Variability of flux density is a key attribute of all Active Galactic Nuclei (AGN) including QSOs. It is an effective tool to probe their spatially unresolved central engines. In the case of radio-loud AGNs, particularly blazars, the optical variability is predominantly associated with the (Doppler boosted) relativistic jet \citep[e.g., see][]{Wagner-Witzel1995ARA&A..33..163W,blandford2019relativistic}. On the other hand, radio-quiet AGNs exhibit milder optical variability on all time scales, which is often thought to arise from perturbations/instabilities in the accretion disk that feeds the central supermassive black hole (SMBH) \citep[e.g.][]{ulrich1997variability,siemiginowska1997deriving,kelly2009variations,bauer2009quasar}, although a weak jet may also be contributing to the variability \citep[e.g., see][]{blundell2001ApJ...562L...5B,GopalKrishna2003ApJ...586L..25G,2004MNRAS.350..175S,barvainis2005radio,Czerny2008MNRAS.386.1557C}. Among the various radio-loud sub-classes of AGNs, BL$~$Lac objects, which are characterized by an essentially featureless optical/UV spectrum, show strong flux variability on time scales ranging from minutes to years \citep[e.g., see][and references therein]{2018BSRSL..87..281G}. Curiously, this property is not shared by a special subclass of radio-quiet AGNs, called `Weak Line quasars' (WLQs), although they too lack strong emission lines in the UV/optical spectrum (the dividing line is conventionally set at the rest-frame equivalent width EW$_r <15$ \AA$~$ for Ly$\alpha+$ N~{\sc v} emission lines). For this and other reasons, WLQs have remained an enigma for the past few decades \cite[e.g., see][]{Fan2006AJ....131.1203F,Anderson2007AJ....133..313A,Plotkin2010ApJ...721..562P,Wu2011ApJ...736...28W,Meusinger2014A&A...568A.114M,shemmer2015weak,ni2018connecting}. In fact, even the possibility has been considered that WLQs might be the elusive radio-quiet analogs of BL Lacs \citep[e.g., see][]{fan1999discovery,Shemmer2009ApJ...696..580S,Plotkin2010ApJ...721..562P}. However, the generally negative searches for strong intranight optical variability (INOV) disfavor this proposal \cite[e.g., see ][]{Gopal2013MNRAS.430.1302G, Chand2014MNRAS.441..726C, Kumar2015MNRAS.448.1463K,kumar2016intranight}, so also  their observed low degree of optical polarisation, $p <3\%$ \cite[e.g., see][]{Smith2007ApJ...663..118S,DiamondStanic2009ApJ...699..782D,Heidt2011A&A...529A.162H,2018MNRAS.479.5075K}.
All this reinforces the notion that WLQs are an intriguing, extreme subset of the normal QSO population, although the relationship between the two remains rather obscure. \par
Proposed explanations for the weak emission lines of WLQs include 
(i) a soft ionizing continuum leading to the emission deficit from the broad-line region (BLR) \citep[][]{Leighly2007ApJS..173....1L,Laor2011MNRAS.417..681L,Wu2011ApJ...736...28W,luo2015x}. Such scenario also includes a geometrically thick accretion disk due to super-Eddington accretion, which creates a gaseous shield around the central continuum source, capable of inhibiting photo-ionization of the BLR clouds \citep[e.g., see ][and references therein]{ni2018connecting,paul2022connecting}. 
and (ii) an anemic, unusual BLR in WLQs \cite[][]{DiamondStanic2009ApJ...699..782D,shemmer2010weak,Nikolajuk2012MNRAS.420.2518N}, perhaps due to their representing an early evolutionary phase of quasars so that the accretion disk is relatively recently established \citep[e.g., see][]{Hryniewicz2010MNRAS.404.2028H,Liu2011ApJ...728L..44L,Meusinger2014A&A...568A.114M} and the BLR region is still underdeveloped \cite[e.g., see][]{Hryniewicz2010MNRAS.404.2028H,andika2020probing,ritish2023evidence}. 
It is clear that the prospects of resolving the puzzle of WLQs rest substantially on comparing them with normal QSOs in terms of additional observables. 
Optical variability on different time scales is one such observable. Comparison of their optical variability on hour-like time scales has been presented in a series of papers \cite[e.g., see ][]{Gopal2013MNRAS.430.1302G, Chand2014MNRAS.441..726C, Kumar2015MNRAS.448.1463K,kumar2016intranight}. However, for both normal QSO and their WLQ subset, such rapid optical variability does not stand out well above the typically achieved detection threshold of 1-2$\%$ (unlike blazars), precluding a quantitative comparison of the two. An alternative is to extend the comparison to much longer (month/year-like) time scales, which is now possible, thanks to the availability of long-term optical light-curves of high sensitivity and cadence, from the Zwicky Transient Facility \citep[ZTF survey\footnote{https://www.ztf.caltech.edu}][]{bellm2018zwicky}. The ZTF survey scans the northern sky in {\it g, r} and {\it i} bands, using a 48-inch Schmidt telescope stationed at Mount Palomar and is equipped with a 47 sq. deg wide-field imager. This database has motivated the present study. Section \ref{sample} describes our sample selection for WLQs and the control sample of normal QSOs. In Section \ref{analysis}, we describe the analysis procedure, followed by presentation of the results and a brief discussion in Section 4. The last section summarises our main conclusions. Throughout, we have assumed the flat Universe with $\text{H}_0$ = 70 $\rm km\ s^{-1}\ Mpc^{-1}$, $\Omega_m$ = 0.3 and $\Omega_\Lambda$= 0.7.

\section{The WLQ Sample}

\label{sample}
Our sample is derived from the compilation of 90 bona fide WLQs used in \citet[][hereafter KCJ23]{ritish2023evidence}, which, in turn, was derived from a parent sample of 108 WLQs based on the compilations by \cite{Plotkin2010ApJ...721..562P} and \cite{Meusinger2014A&A...568A.114M}, after excluding sources showing a galaxy-like spectrum, or a significant proper motion (for detail, see KCJ23). This sample was supplemented with 13 WLQs from \cite{luo2015x}, 14 WLQs from \cite{ni2018connecting}, and 1 WLQ from \cite{meusinger2022gaia} (the remaining WLQs listed in these publications are already included in the above-mentioned set of 108 WLQs). This sample containing 118 WLQs was then checked for radio loudness, by estimating the radio-loudness parameter R\footnote{Radio loudness (R) of a QSO is usually characterised in terms of the ratio of its flux densities at 5 GHz and at 2500~\AA~in the rest frame, R being  $>$ 10 and $<$ 10 for radio-loud and radio-quiet quasars, respectively \citep[e.g., see][]{Kellermann1989AJ.....98.1195K}.} for each WLQ after cross-matching it to within 5" with the 1.4 GHz catalogues from the Faint Images of the Radio Sky at Twenty-cm (FIRST; \citet{becker1995ApJ...450..559B}. We found 7 of the WLQs to have R $>$ 10 and discarding these radio-loud cases reduced the sample to 111 WLQs. Moreover, we intended to perform the Structure Function (SF) analysis for both {\it g} and {\it r} bands, and the requirement of availability of the ZTF-DR22 light-curves for both these bands was satisfied by only 88 of the WLQs. In order to compare these 88 WLQs with normal QSOs in terms of optical variability, we next assembled from the SDSS DR14 quasar catalog \citep[][]{paris2018sloan}, a preliminary `control sample' of normal QSOs matched in the redshift-magnitude plane, allowing tolerances of 0.01 in redshift and 0.6-mag in the $r$-band apparent magnitude. For each WLQ, these tolerances allowed selection of 10 normal QSOs into the control sample, except for 12 WLQs having very low or very high redshifts ($z$ $<$ 0.06, or $>$ 3.5). The required exclusion of these 12 WLQs led to our final sample of 76 WLQs and the corresponding control sample comprising 10 normal QSOs for each WLQ. A total of 102 out of these 760 normal QSOs had to be discarded as their ZTF $g$- and $r$-band light-curves failed to meet our selection criterion of minimum of 100 data points per light-curve (which was also followed for our WLQ sample selection) and another 55 QSOs had to be excluded for being radio-loud (R $>$ 10). Thus, we were finally left with a control sample of 603 normal QSO, for our sample of 76 WLQs (see Table~\ref{wlq_sample}).
\section{Analysis}
\label{analysis}
\begin{figure}
\centering
    \includegraphics[width = 0.4\textwidth,height=0.25\textwidth]{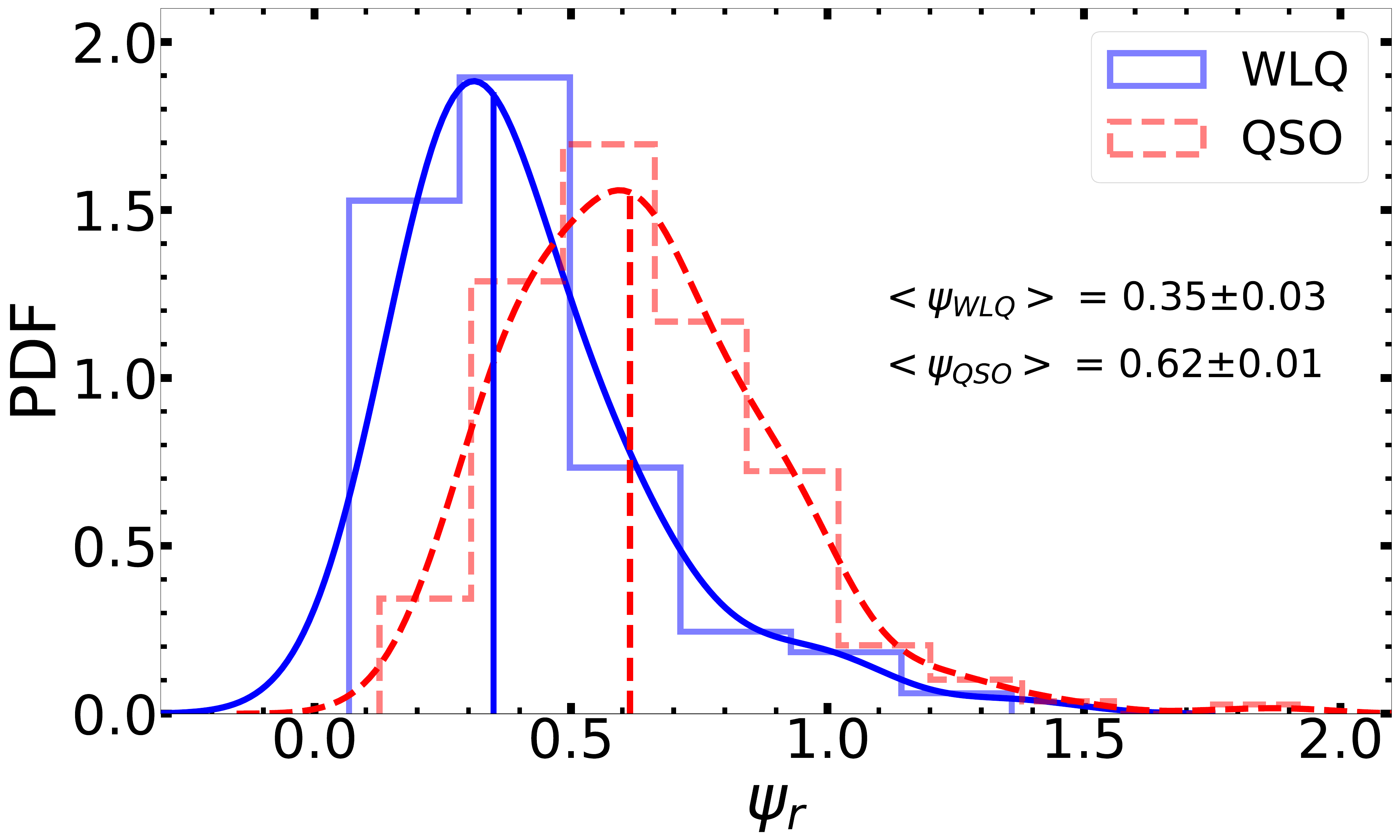}
    \includegraphics[width = 0.4\textwidth,height=0.25\textwidth]{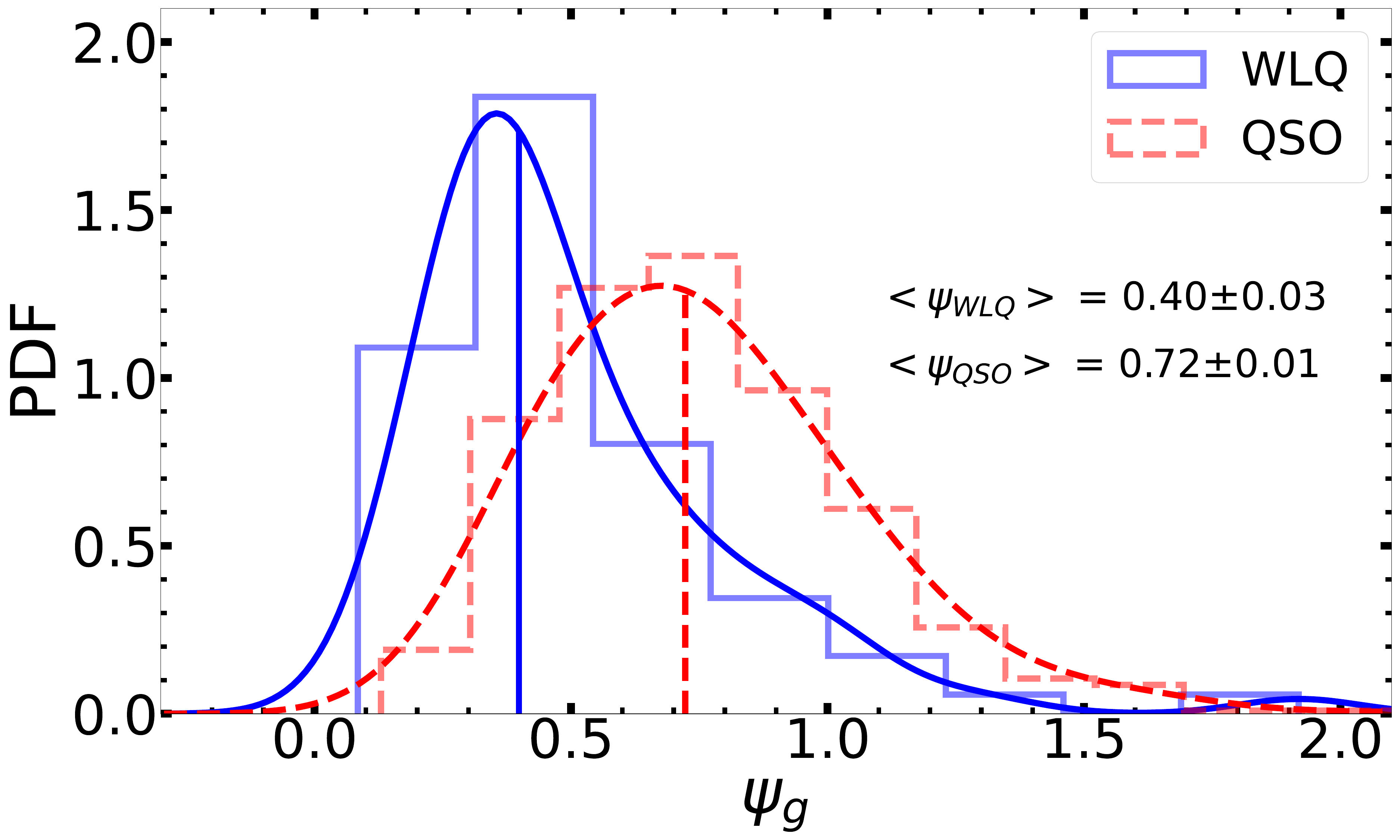}
\caption{The kernel smooth probability distribution function (PDF) of the variability amplitude ($\psi$) for the $r$-band (upper panel) and $g$-band (lower panel) light-curves of the WLQ sample (blue solid line) and the matched control sample of normal QSOs (red dashed line). The median values of $\psi$ are also shown. The distributions of $\psi$ in the $g$-band and $r$-band among the WLQs and the normal QSOs are  significantly different, with the KS-test 
P$_{null}$ of 2.68$\times~10^{-15}$ and 4.34$\times~10^{-14}$, respectively.}
\label{fig:flux_var_all}
\end{figure}
We have used the ZTF light-curves to determine the distributions of variability amplitude ($\psi$) for our WLQ sample and the control sample of normal QSOs. 
For each light-curve, $\psi$ was computed using the relation \citep{Heidt1996A&A...305...42H}: 
\vspace{0.2cm}
\begin{equation}
    \psi = \sqrt{(A_{max}-A_{min})^2-2\sigma^2}
\label{psi_r}
\end{equation}
Here $\sigma^2 = <\sigma_{i}^2>$, with $\sigma_{i}$ being the rms photometric uncertainty for the $i^{th}$ data point in the light-curve. $A_{max}$ and $A_{min}$ are, respectively, the maximum and minimum amplitudes in the light-curve. The distributions of $\psi$ for the two optical bands are shown in Fig. \ref{fig:flux_var_all}
for the WLQ sample and the control sample of normal QSOs. The corresponding median values of $\psi$ are 0.35$\pm$0.03 \& 0.62$\pm$0.01 for the $r$-band, respectively and 0.40$\pm$0.04 $\&$ 0.72$\pm$0.01 for the $g$-band. Thus, based on the ZTF light-curves spanning $\sim$ 6 years, it is evident that,  compared to normal QSOs, WLQs exhibit significantly milder variability, in both bands. Furthermore, variability amplitude appears to be greater in the $g$-band, consistent with the well-known trend for the optical variability of AGN to increase towards higher frequencies \citep{hawkins2002variability,wilhite2008variability,sun2014discovery}.  
Below, we present Structure-Function analysis to investigate the time scales associated with the weaker optical variability of WLQs, compared to normal QSOs. \par
The Structure Function quantifies variability in a data sequence by measuring the magnitude difference between two observations in a light curve,  separated by a chosen time lag, \(\Delta t\), defined in the emitter's rest-frame. The strength of SF analysis lies in the mutual independence of the different bins of \(\Delta t\) \citep[e.g.,][]{de2005structure,de2022structure}. Such a statistical approach can be readily applied to a large sample of objects, for quantitatively estimating their `ensemble' variability over a range of (intrinsic) time scales. Various definitions of SF exist in the literature \citep[e.g.,][]{hawkins2002variability,de2005structure,Meusinger2011A&A...525A..37M,graham2014novel,Kozlowski2017ApJ...847..144K}.  Here, we shall compute the modified 
SF, defined as (\cite{di1995variability}):  
\begin{equation}
SF(\Delta t) = \sqrt{\frac{\pi}{2} \langle |m(t+\Delta t) - m(t)| \rangle^2 - \langle \sigma^2 \rangle}
\end{equation}  
Here, \(m(t)\) and \(m(t+\Delta t)\) are the apparent magnitudes at times \(t\) and \(t+\Delta t\) within a QSO light-curve, with t defined in the QSO's rest-frame. The term \(\langle |m(t+\Delta t) - m(t)| \rangle\) represents the average of the absolute values of the magnitude differences for all pairs of data points in the entire set of light-curves of a QSO sample, where each pair is separated by a chosen time lag \(\Delta t\) defined in the rest-frame of the corresponding QSO.  The error $\sigma$ on the magnitude difference for a pair is the square root of the quadratic sum of the rms errors of the two magnitude values (assuming a Gaussian-error distribution). The term \(\langle \sigma^2 \rangle\) represents the average of $\sigma^2$, taken over all the pairs in the chosen bin of rest-frame time lag \(\Delta t\). The factor \(\frac{\pi}{2}\) accounts for the assumption that both intrinsic variability and the noise on individual data points follow Gaussian distributions \citep{wilhite2008variability}. The statistical uncertainty for each point of the computed `ensemble' SF is estimated by properly propagating the computed errors on the magnitude differences for all the pairs within the respective bin of  \(\Delta t\). The derived ensemble SFs for our WLQ sample and for the control sample of normal QSOs (both radio-quiet) are displayed in Fig. \ref{fig:sf_3a} and \ref{fig:sf_3b}, for the $r-$band and $g-$band light-curves. Note that the error bars on the symbols plotted are smaller than their sizes. As an independent check on these ZTF-based ensemble SFs, we have repeated the SF computation using the light-curves taken from the Catalina Real-Time Transient Survey (CRTS) \citep{Drake_crts2009ApJ...696..870D}. Note that only 69 out of the 76 WLQs in our ZTF sample fall within the CRTS region. Also, with a limiting magnitude V $\sim$ 20 in a single exposure, CRTS is less sensitive compared to the ZTF survey for which the limiting magnitude is $\sim$ 21 ($g-$band). To make an allowance for this, we have reset the magnitude limit at 1-mag brighter level, for defining the WLQ sample from the CRTS. The sample size available for computing the SF thus got reduced to 51 WLQs (all radio-quiet). Fig. \ref{fig:crts_sf} shows the SF (V-band) computed for this WLQ sample, together with the SF (V-band) computed for its control sample which consists of 361 normal (radio-quiet) QSOs, selected in the same manner as for our ZTF-based analysis.\par
{\renewcommand\thefigure{2a} 
\begin{figure}
    \centering
    \includegraphics[width=0.5\textwidth,height=0.33\textwidth]{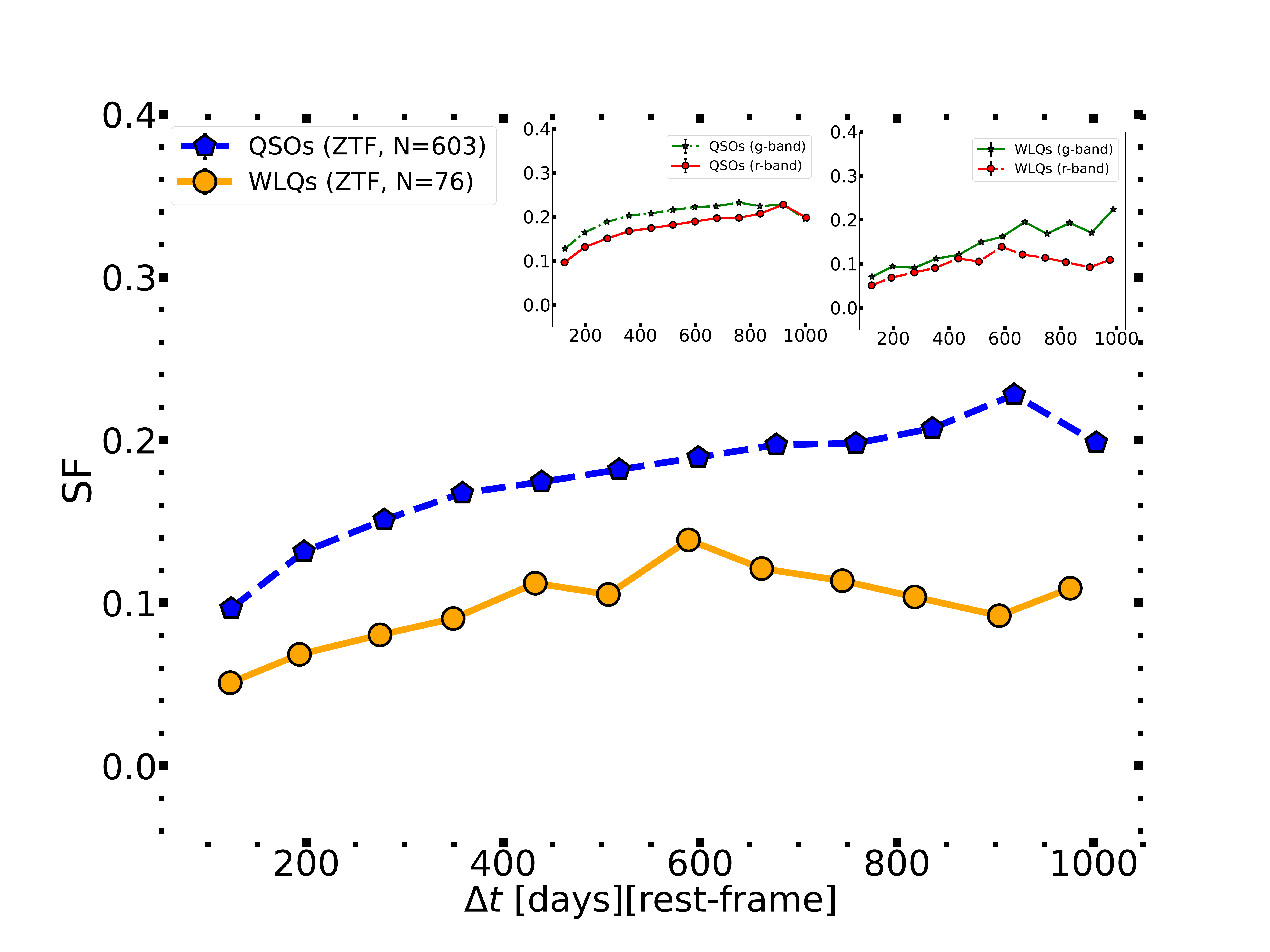}
    \caption{The computed ensemble Structure Function (SF) for the WLQ sample (orange) and for the control sample of normal QSOs (blue), derived using their ZTF $r$-band light-curves. The error bar on each plotted symbol is smaller than the symbol size. The two insets compare the computed SFs in $g$-band and $r$-band for the WLQs and normal QSOs.}
    \label{fig:sf_3a}
\end{figure}
}
{\renewcommand\thefigure{2b} 
\begin{figure}
    \centering
    \includegraphics[width=0.5\textwidth,height=0.35\textwidth]{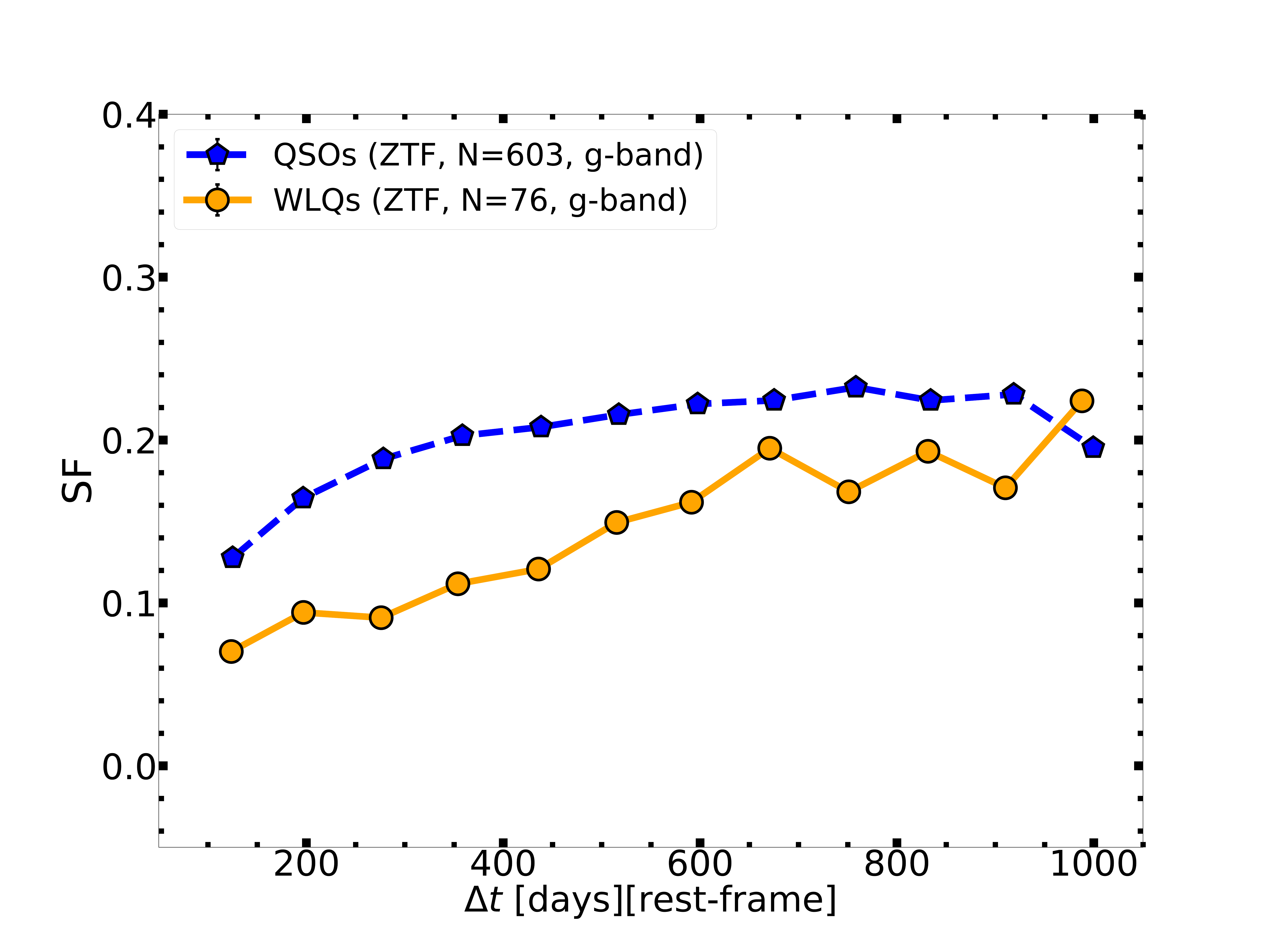}
    \caption{The computed ensemble Structure-function (SF) for the WLQ sample (orange) and for the control sample of normal QSOs (blue), derived using their ZTF $g$-band light-curves. The error bar on each plotted symbol is smaller than the symbol size.}
    \label{fig:sf_3b}
\end{figure}
}
{\renewcommand\thefigure{3}
\begin{figure}
\centering
    \includegraphics[width = 0.50\textwidth,height=0.32\textwidth]{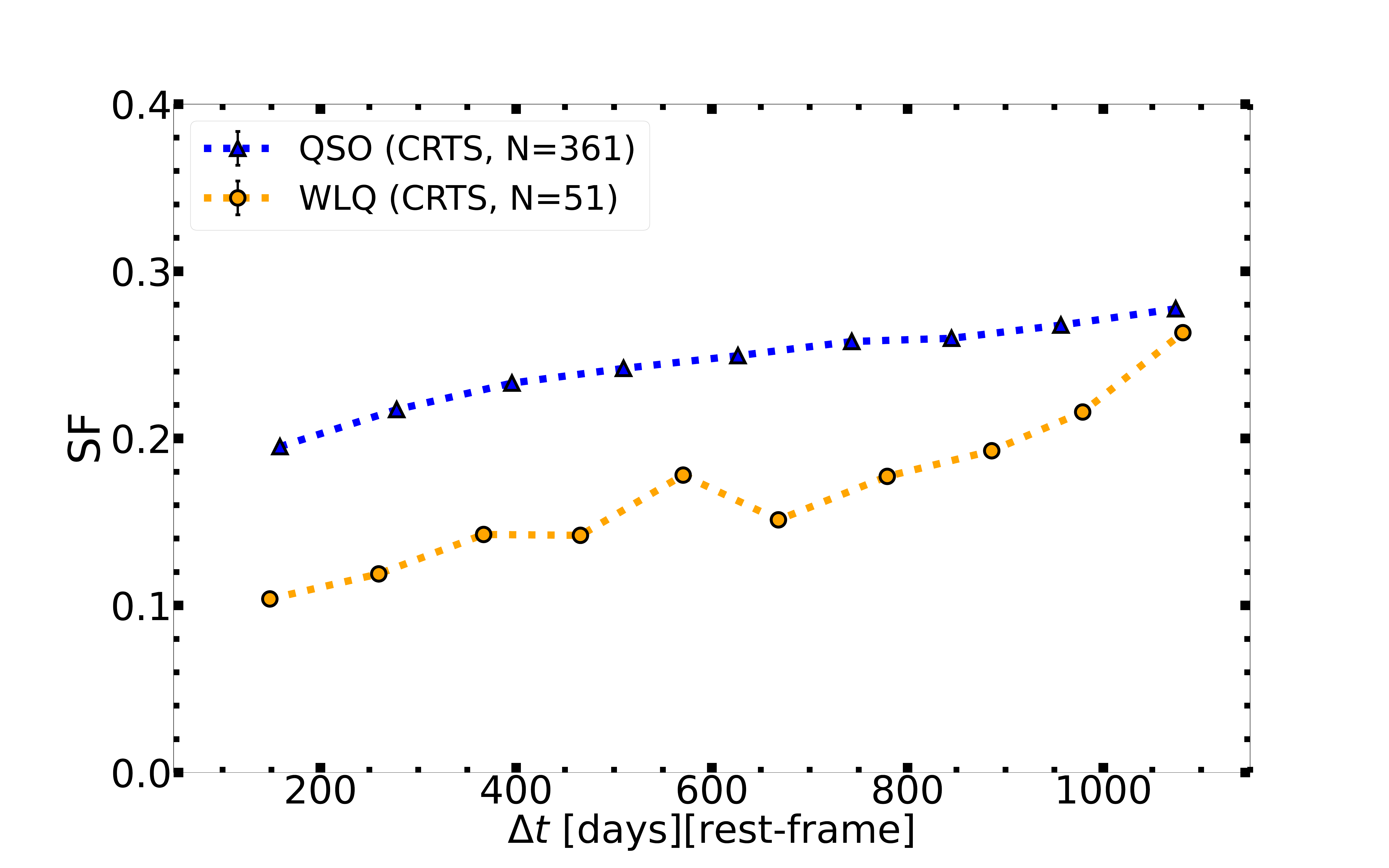}
\caption{The computed ensemble Structure-function (SF) for the sample of 51 WLQs (orange) and for the control sample 361 normal QSOs (blue), based on the V-band light-curves taken from the CRTS survey. The error bar on each point is smaller than the symbol size.}
\label{fig:crts_sf}
\end{figure}
}
\section{Results and Discussion}
\label{discussion}
Fig. \ref{fig:sf_3a}, \ref{fig:sf_3b} and Fig. \ref{fig:crts_sf} display the most salient and statistically robust findings from the present work based on the ZTF and CRTS surveys. The computed ensemble SFs for the light-curves drawn from these two surveys consistently show that on month/year-like time scales, WLQs exhibit systematically milder optical variability, compared to normal QSOs (recall that the members of both the WLQ sample and the control sample of normal QSOs are radio-quiet, minimizing any contamination from jet-related optical continuum). The ensemble variability amplitudes for the two QSO types are thus found to differ typically by a factor of $\sim$ 1.76$\pm$0.05, for the month/year-like time scales probed here. Although a hint for such a difference was present in an earlier study, based on a much smaller sample of WLQs with their CRTS (V-band) light-curves then available for just a 3-year time-span \citep[see][]{2018MNRAS.479.5075K}, the difference between the optical variability levels of WLQs and normal QSOs is recognized, for the first time, in the present study. \par
The slopes of the SFs (ZTF) of the WLQs and of their control sample of normal QSOs have been determined by straight-line fitting to the SFs on the log-log scale. This gave power-law slopes of 0.55$\pm$0.05 $\&$ 0.41$\pm$0.04 in the $r$-band and 0.51$\pm$0.06 $\&$ 0.34$\pm$0.04 in the $g$-band, for the WLQs and normal QSOs, respectively (Fig. \ref{fig:sf_3a}, \ref{fig:sf_3b}). The corresponding slopes for the SF (CTRS) are found to be 0.41$\pm$0.06 and 0.18$\pm$0.01, respectively (Fig. \ref{fig:crts_sf}). These estimates are broadly consistent with the power-law slopes of $0.36 \pm 0.01$ \citep{bauer2009quasar} and 0.246$\pm$0.008 \citep{vanden2004ApJ...601..692V}, estimated for normal QSOs. Note that a SF slope of $\sim$ 0.44 has been theoretically predicted for the accretion disk instability model of flux variability \citep{hawkins2002variability}.\par
It is relevant to consider the possibility that 
in our redshift and $r$-band magnitude matching (see Sect. \ref{sample}), the flux contributed by any emission line falling within the $r$-band, depending on source redshift, would be expectedly higher for normal QSOs in comparison to the WLQs which have distinctly weak or absent emission lines. As a result,  the control sample would be systematically less luminous than the WLQ sample, in terms of (line-less) continuum luminosity. Now, the net contribution of the emission lines in a standard QSO spectrum at some reference redshift, to the observed photometric band would vary as the spectrum sweeps past the band with changing redshift. This variation is reflected in the published $K$-correction values (defined as $m_{intrinsic} = m_{observed} - K(z)$), which also include the redshift-dependent differential contribution coming from the underlying continuum falling within the observed photometric band. Thus, the maximum range observed in the $K(z)$ values amounts to an upper limit to the flux contributed by any emission lines falling within the observed photometric band. Therefore, for a conservative estimate of the expected emission-line contribution to the observed luminosity 
we draw a histogram of the K-correction values for a normal QSO \citep[Table 4 of][]{richards_2006AJ....131.2766R}, covering the range 0.5 $<$ $z$ $<$ 3.5 which essentially overlaps the $z$-range for our control sample of normal QSOs. \par
As seen from Fig. \ref{fig:k_corr_dist}, the range spanned by the K-correction values ($-$0.4-mag to $+$0.2-mag) amounts to a maximum change $\Delta m = 0.6$-mag \footnote{The K-corrections given by these authors are actually for the $i-$band, but the 0.6-magnitude range should be a fairly good approximation for the $r$-band.}, implying that emission lines could have brightened the normal QSO (having strong emission lines) by at most 0.6-mag, vis-à-vis the WLQ.
Thus, the optical continuum luminosity of our WLQs sample could be systematically higher than the value for the QSO control sample, by a factor of 1.7, at the most. Could this luminosity mismatch account for the milder optical variability observed for the WLQs, as compared to the normal QSOs, given that optical variability is known to anti-correlate with optical luminosity \citep[e.g., ][]{MacLeod_2010ApJ...721.1014M,caplar2017optical,laurenti2020individual}? To check this we use the analytical fit to the anti-correlation \citep[Eq. 9 of][]{laurenti2020individual}, from which it follows that in order to account for the 1.76$\pm$0.05 times milder variability of the WLQs (see above), the WLQ sample would have to be systematically $\sim 20$ times more luminous in the optical than the control sample of normal QSOs. This requirement is an order-of-magnitude larger than the estimated maximum brightening of a normal QSO, at any $z$ within the range considered here, due to the contribution from emission-lines within the $r$-band (see above).
Therefore, the distinctly milder optical variability of WLQs, as found here, is extremely unlikely to be an artifact resulting from the inability to remove the emission-line contribution from the published magnitudes of each QSO, particularly the normal QSOs which constitute the control sample. \par
Furthermore, at least qualitatively, the observed milder optical variability of the WLQs (which are thought to be accreting at a higher Eddington rate, see Sect. \ref{introduction}), seems consistent with the observed anti-correlation between optical variability and (dimension-less) accretion rate \citep[][]{Lu_2019ApJ...877...23L}. From a theoretical perspective, it is tempting to explore the possibility of a common physical cause underlying the abnormally weak broad emission-lines in WLQs and their milder optical continuum variability, in comparison to the general QSO population. Conceivably, both these exceptionalities of WLQs, as found here, may be traceable to the physical conditions near the inner edge of the dusty torus which not only feeds the accretion disk, but also whose part that extends inside the dust sublimation radius can be identified as the BLR clouds \citep[][]{Antonucci2023Galaxy}. For a given amount of material flowing from the torus into the BLR, if the degree of clumpiness is higher \citep[e.g., see][]{wang2017tidally}, the broad-line emissivity would clearly be enhanced. At the same time, the optical continuum which arises mainly from the accretion disk, may also exhibit stronger variability. The latter expectation derives from the theoretical model according to which a (variable) optical/UV light-curve arises mainly from cascades of expanding optically-thick shocks between colliding clumps of matter flowing in the accretion disk \citep[][]{turler2005cascades,ishibashi2009agn}. \footnote{This AGN accretion disk model, an alternative to the standard `geometrically thin, optically thick accretion disk’ paradigm, was inspired by the observed quasi-simultaneity of the variable optical and UV light-curves \citep[e.g., see][]{courvoisier1991observational,collin1991origin,alloin1985recent}, as well as the independent argument by \citet{antonucci1989lyman} based on the optical polarisation vector of jetted AGN.} Conversely, if the material flowing inward from the inner edge (i.e., the dust sublimation radius) of the torus is more homogenous (i.e., less clumpy), one would expect not only a weaker line emission from the BLR, the defining characteristic of WLQs, but also a less variable optical continuum from the accretion disk, as indeed found here using the extensive sets of well-sampled optical light-curves. It would be interesting to explore in detail this simplified proposal in order to visualize the physical process underlying the WLQ phenomenon. \par
\section{Conclusions}
In an attempt to find additional observational clues for understanding the various peculiarities exhibited by `weak-line QSOs' (WLQs) in the optical/UV/X-ray bands, we have found a new observational differentiator between WLQs and normal QSOs,
namely the optical variability on month/year-like time scales. We assembled a well-defined sample of 76 WLQs and a control sample of 603 normal QSOs (both radio-quiet), via matching the two in the redshift-magnitude plane. Using the $g-$ and $r$-band light-curves for these two samples, recorded in the recently concluded major optical survey ZTF with a typical cadence of 3 days, we have shown that as a class, 
WLQs exhibit distinctly milder optical variability, as compared to the normal QSOs, on month/year-like time scales. Based on a Structure-Function analysis of the ZTF light-curves for these two matched samples of QSOs, the `ensemble' optical variability of WLQs is estimated to be typically lower by a factor of $\sim$ 1.7 in amplitude, on month/year-like time scales. An independent confirmation to this finding has come from the use of the V-band light-curves for a well-defined sample of 51 WLQs and its control sample of 362 normal QSOs, taken from the CRTS survey database. This result has motivated  a simplified, basic scenario linking the abnormal weakness of the broad emission lines of WLQs and their milder optical flux variability (compared to normal QSOs), to a common physical cause, namely the clumpiness of the torus material accreting into the central engine and its immediate environment. Further exploration of this newly found observational clue to the WLQ phenomenon, can be expected to play a significant role in understanding the working of this intriguing subclass of Active Galactic Nuclei. 
\section*{Acknowledgements}
{We thank the anonymous referee for her/his helpful comments on the manuscripts.}
HC and RK are grateful to IUCAA for the hospitality and HPC facility under IUCAA Associate Programme. GK thanks Indian National Science Academy for a Senior Scientist position during which part of this work was done. 

\section*{Data Availability}

The data used in this study are publicly available in the CRTS, ZTF  DR16, and SDSS DR16 Data Release.



\bibliographystyle{mnras}
\bibliography{references} 



\appendix
\section{Some Extra Material}

\begin{figure}
\centering
    \includegraphics[width = 0.45\textwidth,height=0.25\textwidth]{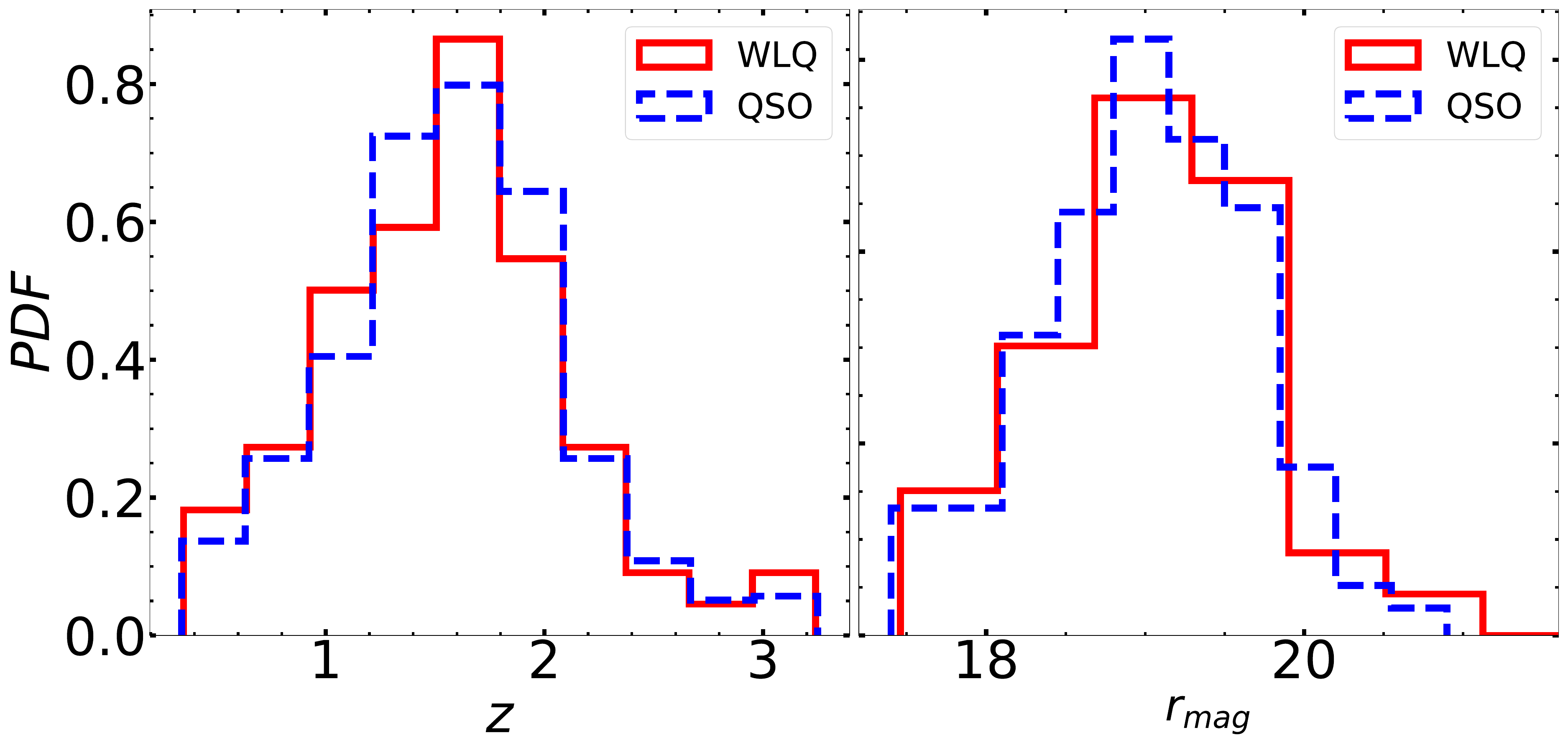}
\caption{The normalized distribution of redshift as well as the $r$-band magnitude for the present sample of 76 WLQs and the control sample consisting of 603 normal QSOs matched in redshift and $r$-band magnitude (Sect. 2). The solid red line represents the WLQ sample whereas the dashed blue line represents the control sample of normal QSOs. The samples of WLQs and normal QSOs are statistically indistinguishable in terms of both redshift and $r$-band magnitude, with KS-test P$_{null}$ of 0.98 and 0.95 respectively.}
\label{fig:redshift_dist}
\end{figure}
\begin{figure}
\centering
    \includegraphics[width = 0.49\textwidth,height=0.30\textwidth]{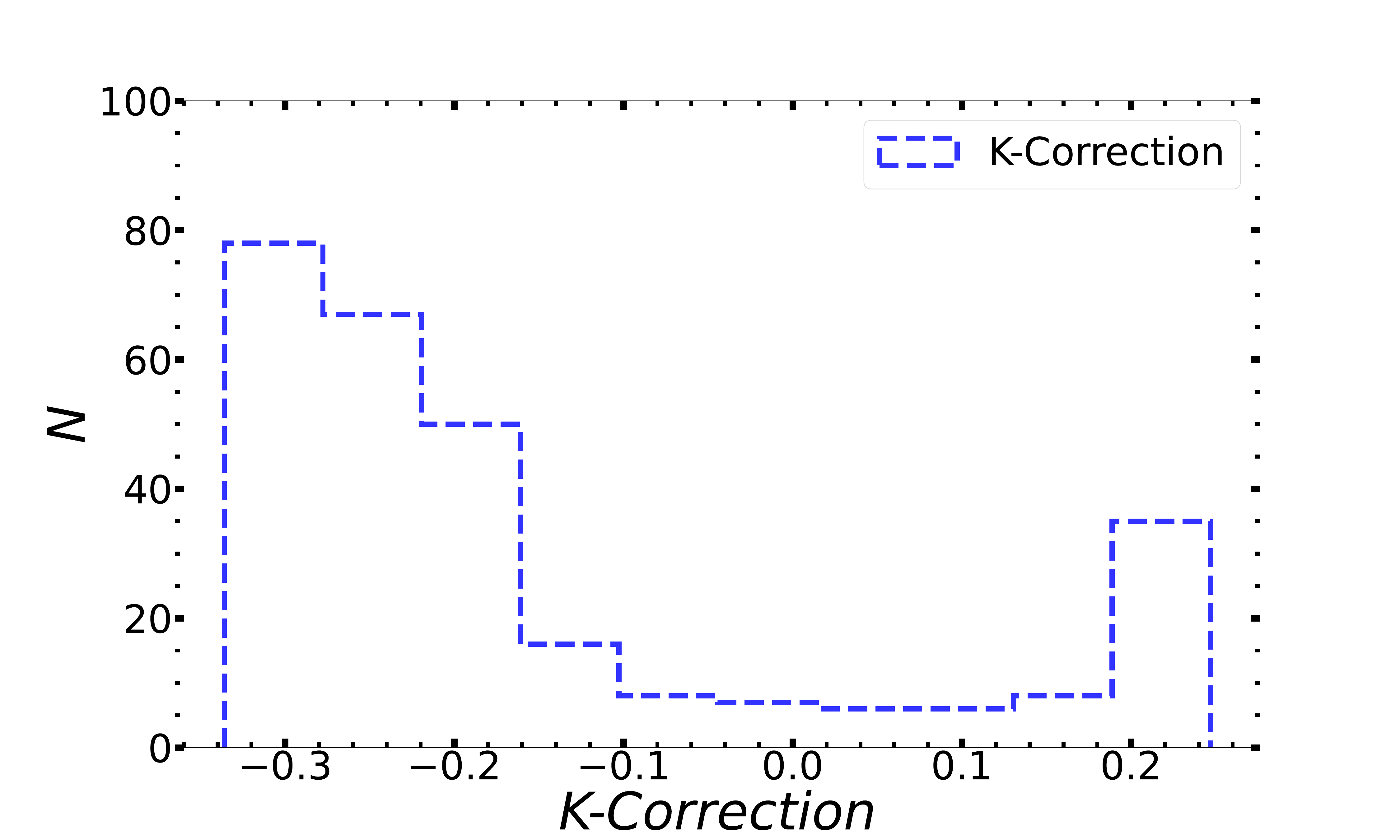}
\caption{The distribution of K-Correction factor for the QSOs taken from Table 4 of \citet[][]{richards_2006AJ....131.2766R} within the redshift range of 0.5$<$z$<$3.5.}
\label{fig:k_corr_dist}
\end{figure}
\begin{table}
	\centering
	\setlength\tabcolsep{1.1pt}
	\caption{ Basic parameters of the 76 WLQs in our sample. }
	\begin{tabular}{ccccc} 
		\hline
		Source Name & RA & DEC & z & $r_{mag}$ \\
		&  (Degree) &   (Degree) &  &  (SDSS) \\
		\hline \\
  \vspace{0.02in}
		J001444.03$-$000018.5 & 3.68 &  $-$0.005 & 1.55  & 18.22$\pm$0.02 \\[0.05in]
            J001514.88$-$103043.6 & 3.81 & $-$10.51  & 1.17  & 19.60$\pm$0.02 \\[0.05in]
            J005713.01$+$004205.7 & 14.304 & 0.702 &  1.54 & 19.88$\pm$0.02 \\[0.05in]
		--- & --- & --- & --- & ---\\
		\hline \\
	\multicolumn{5}{{p{\columnwidth}}}{\textbf{Note:} The entire table is available in online version. Only a portion of this table is shown here to display its form and content.}
	\end{tabular}
 \label{wlq_sample}
\end{table}
\begin{table}
	\centering
	\setlength\tabcolsep{1.1pt}
	\caption{ Basic parameters of the 603 QSOs in the control sample. }
	\begin{tabular}{ccccc} 
		\hline
		Source Name & RA & DEC & z & $r_{mag}$ \\
		&  (Degree) &   (Degree) &   &  (SDSS)  \\
		\hline \\
  \vspace{0.02in}
            J000139.21$+$204000.2 & 0.413 & 20.666 & 1.542  & 18.19$\pm$0.02\\[0.05in]
            J000155.68$-$073508.6 & 0.482 & $-$7.5857 & 2.756  &19.54$\pm$0.02 \\[0.05in]
            J001223.31$-$010922.7 & 3.097 & $-$1.156 & 1.4305 &19.47$\pm$0.03 \\[0.05in]
		--- & --- & --- & --- & ---\\
		\hline \\
	\multicolumn{5}{{p{\columnwidth}}}{\textbf{Note:} The entire table is available in online version. Only a portion of this table is shown here to display its form and content.}
	\end{tabular}
 \label{qso_sample}
\end{table}

\bsp	
\label{lastpage}
\end{document}